\documentclass[11pt]{article}
\usepackage{epsfig,geometry} 
\title{Quasispecies and recombination}
\author{Martin Nilsson Jacobi\footnote{{\tt mjacobi@chalmers.se}} and Mats Nordahl \\
		Chalmers University of Technology\\
		Gothenburg, Sweden.}

\begin{document}

\maketitle

\begin{abstract}

Recombination is introduced into Eigen's
theory of quasispecies evolution. Comparing numerical simulations 
of the rate equations in the
non-recombining and recombining cases show that 
recombination has a strong
effect on the error threshold and, for a wide range of mutation rates,
gives rise to two stable fixed points in the dynamics. This bi-stability
results in the existence of two error thresholds. We prove that, 
under some assumptions on the fitness landscape but for general crossover probability, 
a fixed point localized about the sequence with superior fitness is globally
stable for low mutation rates.  

\end{abstract}

\section{Introduction}

\label{introduction}

The quasispecies concept was introduced by Eigen in 1971~\cite{Eigen71}
to describe populations of self-replicating molecules.
A quasispecies is an equilibrium distribution of closely related gene 
sequences, localized in sequence space around one or a few sequences 
of high fitness. The quasispecies model can be viewed as a simple
framework that contains all the basic ingredients of Darwinian evolution. 
In particular, it captures the critical relation
between mutation rate and information transmission~\cite{Eigen71,Eigen77}.
The behavior of these systems has been extensively studied, 
see for instance~\cite{Eigen71,Eigen77,Schuster86,Schuster85,Swetina88}. 
Quasispecies have also been fruitfully studied using concepts and
techniques from statistical physics, see, e.g.,
\cite{Leuthausser86,Tarazona92,AF98}.

In the quasispecies model, the population dynamics is described
on the gene level, and a fitness landscape~\cite{Wright} is used to 
define the degree of adaptation directly from the gene sequence. 
Considerable amounts of work has gone into defining models of
rugged landscapes and analyzing their consequences for the
evolutionary dynamics (e.g.~\cite{Kauffman87,Palmer91,Fontana93,Macken91,Stadler95a}).

In this paper we introduce recombination into the quasispecies
model. With some exceptions (see, e.g., \cite{Boerlijst,OH98,Stadler96,Feldman}) 
previous work on quasispecies has only considered 
non-recombining populations where variation is created only by
mutation. However, most species in nature use crossover during replication, at least to some degree, which makes
this an important case to study. 
Besides applications to evolutionary biology,
developing an understanding for the dynamics of systems under
recombination is also important for gaining theoretical
insights into the behavior of genetic algorithms \cite{Holland75} in 
combinatorial optimization problems. 

Recombination introduces a non-linearity in the rate equations,
which in general results in the appearance of two stable fixed points.
For a wide range of mutation rates this divides the space of initial 
distributions into two regions: one where the population converges to
a distribution localized around the 
genome with highest fitness, and another where it converges to 
an approximately uniform distribution. 
This behavior is qualitatively different from that 
of non-recombining populations. Another interesting observation 
is the shift in the error threshold. 

The main contribution of the paper is a proof that, for a class of
fitness landscapes (see Section~\ref{singlefix} for details),
independent of the crossover probability, there exist exactly 
one globally stabile fixed point. The single peaked fitness landscape
is a special case that belongs to this class.

The rest of this paper is organized as follows:

Section~\ref{quasi} gives a short review of quasispecies evolving 
under mutation only, for comparison with the recombination case.
In section~\ref{recombination}, we introduce the rate equations for quasispecies
with mutation and recombination, and formulate a condition for 
the equilibrium distribution
as a generalized non-linear eigenvalue problem.
 
Section~\ref{num} contains results from numerical simulations of the rate 
equations for a recombining population. We demonstrate how the equilibrium distribution changes
with mutation rate for different initial distributions. 
As in the non-recombining case, a phase transition from a localized to
a uniform distribution occurs
when the mutation rate is increased. The dependence of the phase
transition point on the initial distribution is investigated.

In section~\ref{singlefix} we prove that, under some assumptions on the fitness landscape
but without constraint on the crossover,
when the mutation rate is low enough all initial distributions converge
to a fixed point localized around the genome with highest fitness. 
Finally, section~\ref{discussion} contains a discussion and conclusions.

\section{Quasispecies }

\label{quasi}

In this section we give a short review of relevant results
for quasispecies with non-recombining replication~\cite{Eigen71,Eigen77},
to allow us to compare with the results when recombination is included.
In the model, a self-replicating molecule is represented by a sequence of  
bases $s_k$, $\left( s_1 s_2 \cdots s_n \right)$. The bases are assumed 
to be binary $\{ 0, 1 \}$, 
and all sequences have equal length $n$. A genome is then
given by a binary string $\left( 011001 \cdots \right)$, which also 
can be represented by an integer $k$ ($0 \leq k < 2^n$). 
The space of all gene sequences in the model is called 
sequence space~\cite{Maynard70}. A quasi-species is defined as a  
distribution of sequences localized in sequence space. 

Selection in the quasispecies
model is expressed in terms of a fitness landscape,
which is a function of the phenotype and the environment.
The environment describes direct interactions with other organisms 
as well as the physical environment. 
In the quasispecies model we assume that the phenotype is directly 
determined by the genotype. There is no direct interaction between 
individuals in the population, only indirect competition for resources.
The fitness landscape can then be expressed as a function of the genotype only.
In the following, we only consider a simple landscape
with a single sequence of high fitness $A_0$, called the master sequence,
and with all other sequences $i$ having equal fitness $A_i < A_0$. 
 
Mutations are described by $Q_k ^l$,
the probability that replication of genome $l$  gives genome
$k$ as offspring. If the mutation rate per base, $p_m= 1 - q$, 
where $q$ is the copying accuracy per base, is assumed to be
constant in time and independent of position in the genome,
we obtain
\begin{eqnarray}
    Q_k ^i & = & p_m ^{h_{k i}} q ^{n - h_{k i}} = q ^n 
    \left( \frac{1-q}{q} \right) ^{h_{k i}} \label{eq1}
\end{eqnarray}
where $h_{k i}$ is the Hamming distance between genomes
$k$ and $i$. 

The rate equations that describe the dynamics of the population
are then given by (where $x_k$ denotes the relative concentration
of species $k$): 
\begin{eqnarray}
     \dot{x} _k & = & \sum _l Q_k ^l A_l x_l - e x_k 
             \label{eq2}
\end{eqnarray}
where $e  =  \sum _l A_l x_l$.
The second term ensures 
the total normalization of the population ($\sum _l x_l = 1$).

These differential equations can be solved analytically~\cite{Jones,Thomson}.
Equation (\ref{eq2}) can be made linear through a change of variables 
and we can then use standard techniques to find $x_k$.  If all the elements
of the matrix $Q_k ^l$ are strictly positive, $x_k$
always converges to a unique stable fixed point~\cite{Bellman},
given by the eigenvector
corresponding to the largest eigenvalue $\l = e$ of the matrix
$Q_k ^l A_l$.

For a landscape where the fitness only depends on the Hamming distance
from the master sequence, we can divide sequence space into error classes 
containing sequences with the same number of ones. 
The effective dimension of the system of equations
(\ref{eq2}) can then be reduced from $2^n$ to $n+1$ by summing over
error classes. In this way we obtain the new equations

\begin{eqnarray}
     \dot{x}_K & = & \sum _L \tilde{Q}_K ^L A_L x_L - E x_K
    \label{eq4}
\end{eqnarray}
where the indices $K$ and $L$ denote error classes, and
$\tilde{Q}_K ^L$ describes mutation probabilities between 
error classes rather than sequences.

\begin{figure}[h]
\centering
\leavevmode
\epsfxsize = .75 \columnwidth
\epsfbox{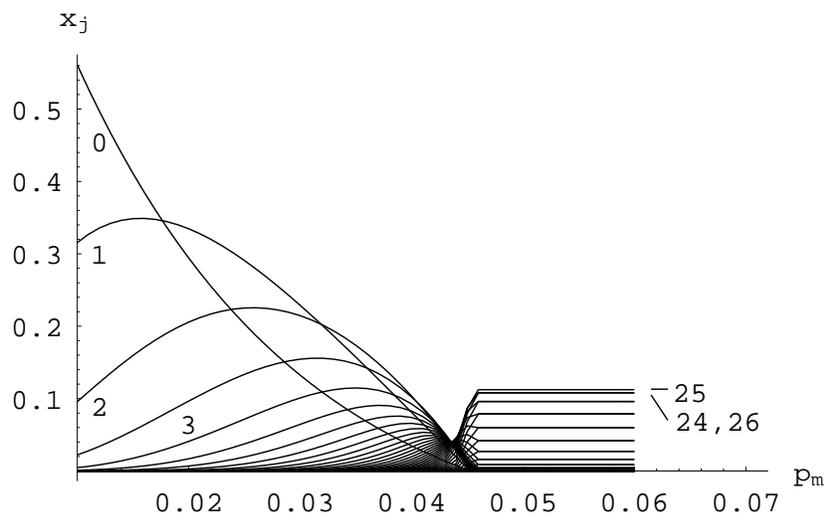}
\caption{The relative equilibrium concentrations of the 51 different error classes
for sequences of length 50 for different mutation rates. The fitness landscape 
has a single peak
$A_0 = 10$, and $A_L = 1$ $ \forall L \neq 0$. The error catastrophe occurs around
$p_m \approx 0.045$.}. 
\label{plotmut50}

\end{figure}

We now consider a fitness landscape with $A_0 = 10$, and
$A_L = 1$ for all $ L \neq 0$. The sequences are indexed by their 
Hamming distance from the master sequence. The equilibrium distributions
corresponding to different mutation rates, $p_m$,
are shown in figure~\ref{plotmut50}. There is a sharp 
transition between a state where the population is localized around
the master sequence $x_0$ and a state where the
distribution is approximately binomial. This is the error
catastrophe (or error threshold) of Eigen and coworkers.\\
\\
Fig.~\ref{plotmut50} here.
\\

The error catastrophe occurs
approximately when $q ^n A _0 / A_i = 1$, or
when the selective advantage of the master sequence, $A_0 / A_i$, is
compensated by the finite probability $q^n < 1$ for the master 
sequence to replicate to itself.  

This observation is important for theories of prebiotic evolution 
of life. When polynucleotides replicate without replicase enzymes, the copying
fidelity is unlikely to exceed 0.99, which means that $n$ cannot be larger than
100~\cite{Eigen71}. This is much smaller than coding regions for replicase enzymes, 
which are needed to increase the copying fidelity. This contradiction is
often called Eigen's paradox. There have been several different attempts to resolve this
problem, such as hyper-cycles~\cite{Eigen77}.

In the following sections we consider quasispecies where both recombination 
and mutation can occur during
replication. The introduction of recombination will cause major changes
in the population dynamics. As an example, we observe that the rate 
equations have multiple stable fixed points. The error threshold also
also significantly shifted.

\section{Recombination}

\label{recombination}

The crossover operator, $T_k ^{l m}$, denotes the
probability that parents $l$ and $m$ give rise to the offspring $k$ in one
recombination event~\cite{Boerlijst,Stadler96}. 
The crossover operator $T_k ^{l m}$ depends on the 
crossover probability $p_c \in [ 0 , 0.5 ]$, i.e., the probability per base pair
for the reading process to switch from one parent to the other.
As an example, $p_c = 0.5$ (uniform crossover) means that each position in the
genome is chosen with equal probability from each parent. Another
extreme case is $p_c = 0$ which means that the offspring inherits all
its genome from a single randomly chosen parent.

The crossover operator has the following properties

\begin{eqnarray}
  &&  0 \leq  T_k ^{l m} \leq 1 \label{eq5} \\
  &&  \sum _k T_k ^{l m} =  1 \:\:\:\:  \forall l,m \label{eq6}  \label{eq7} 
\end{eqnarray}
For uniform crossover we can write $T_k^{l m}$ explicitly as 
\begin{eqnarray}
        T_k^{l m} & = & \left\{ \begin{array}{lcl} 2^{- h_{l m}} & \mbox{if} & O(k,l,m) = 1 \\
                                                0 & \mbox{if} & O(k,l,m) = 0 \end{array} \right.
\end{eqnarray}
where $O(k,l,m) = 1$ if at each position where the parents genome $l$ and $m$ are identical,
the same base also appears in the child genome $k$, else $O(k,l,m) = 0$. New genes can only be created by mutations.

The most realistic and interesting population dynamics involves both recombination
and mutations. In our model we have only recombining individuals and the point
mutations will come in as limited reading accuracy in the crossover process. We have
 chosen to let the number of offsprings depend on both parents. 
The rate equations for a population of sequences which both recombine
and mutate are then given by

\begin{eqnarray}
     \dot{x}_k & = & \sum _{l m} V _k ^{l m} A_l x_l A_m x_m - c x_k \label{eq8}
\end{eqnarray}
where $V _K ^{l m} = \sum _i Q_k ^i T_i ^{l m}$ and $c = \left( \sum _l A_l x_l\right)^2$ (which
is used to normalize the total growth as before).

The rate equations in the case of recombination are in general much harder
to analyze than in the case of pure mutations.  The crossover
operator acts on pairs of sequences, which gives rise to a non-linearity in 
the growth term. We are mainly interested
in the equilibrium distribution, i.e., the concentration of sequences after long time.
In the pure mutation case the stable equilibrium distribution could be calculated
 by solving a standard eigenvalue problem. When recombination is used
the fixed points of the rate equations (\ref{eq8}), $\vec{y}$, are solutions to
the generalized eigenvalue problem:

\begin{eqnarray}
    \sum _{l m} V _k ^{l m} A_l y_l A_m y_m & = & \lambda y_k \;\;\; \forall k \label{eq9}
\end{eqnarray}

All normalized ($\sum _l y_l = 1$) solutions 
to (\ref{eq9})  are also fixed points to the rate equations, since summing over $k$ gives the 
relation $\lambda = \left( \sum _l A_l y_l \right) ^2 = c$. There may however exist solutions to equation
(\ref{eq9}) which cannot be normalized to a vector of concentrations, since all elements
must be non-negative.

In general there exists more than one solution to (\ref{eq9}) which can be normalized
to a concentration vector. It turns out that these multiple fixed points can be stable,
see section~\ref{num}.
One of the most important differences between the non-recombining and the recombining case is 
in fact the uniqueness of the equilibrium distribution. 
As we will see in section~\ref{num} the equilibrium distribution of the rate 
equation (\ref{eq8})
depends on the initial distribution
(as was previously observed in other models, e.g.~\cite{Feldman}). 
This behavior is very different 
from the pure mutation
case, where all initial distributions converge to a unique stable fixed point,
as discussed in section~\ref{quasi}.

However, in Section~\ref{singlefix} we present a proof that in the zero
mutation rate limit, the only globally stabile fixedpoint corresponds
to a population totally localized on the fitness peak.

\begin{figure}[h]
\centering
\leavevmode
\epsfxsize = .75 \columnwidth
\epsfbox{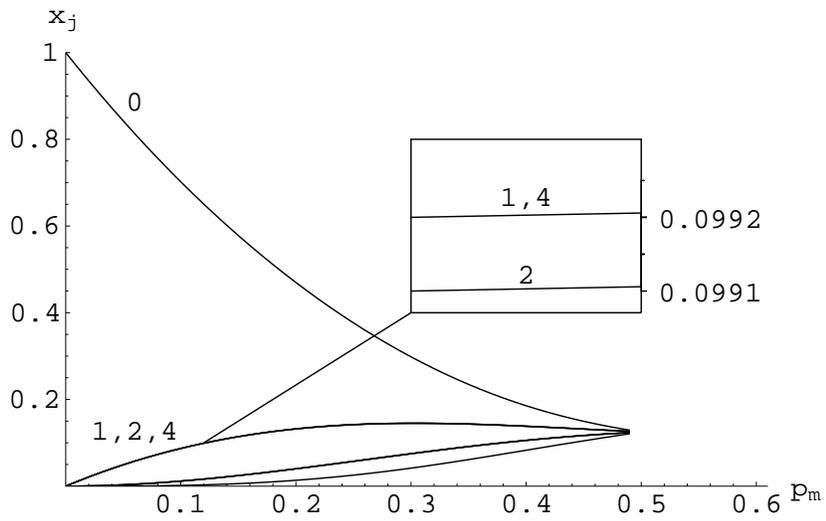}

\caption{The equilibrium distribution for the concentration of genomes
at different mutation rates. The genomes have 
length 4 and the crossover probability $p_c$ is $0.1$. There is a small difference 
in concentration between genomes in the same error class. Genomes
1 and 4 have the same concentration due to the mirror symmetry in the binary strings. 
The symmetry breaking tends to increase with genome length.}
\label{brokensym}
\end{figure}

The dimension of  sequence space scales exponentially with the number of bases
in the genome.  In the non-recombining case we saw
 that the degrees
of freedom in the rate equations (\ref{eq8}) could be reduced from $2^n$ to $n+1$
by dividing the sequences into
error classes. This symmetry is in general broken by recombination (see
figure~\ref{brokensym}). 
The only  non trivial case when the rate equation
(\ref{eq8}) preserves the symmetry between the error classes,  is when $p_c = 0.5$
(uniform crossover). In this case we can write the reduced rate equations as

\begin{eqnarray}
     \dot{x}_K & = & \sum _{L,\:M} \tilde{V} _K ^{L M} A_L x_L A_M x_M - C x_K \label{eq18}
\end{eqnarray}
where we use the same notation as in equation (\ref{eq4}).
For $p_c=0.5$ and $p_m = 0$ the transition probabilities between error-classes 
$\tilde{V}_K ^{LM}$ are given by 
\begin{eqnarray}
      \tilde{V}_K^{L M} & = & \frac{\sum_{d=|M - L|}^{M+L+2\min(n-L-M,0)}
       \left( \begin{array}{c} L \\ 
                \min(l,m) - \frac{2 d - | l-m |}{2} 
                \end{array}\right)}
        {\left( \begin{array}{c} n \\ M \end{array} \right)}
\end{eqnarray}

In the more realistic case when $p_c < 0.5$, we either have to be satisfied with rather
small genome sizes or need to use some approximation method.\\
\\
Fig.~\ref{brokensym} here.
\\

\section{Numerical Results}
\label{num}

\begin{figure}
\centering
\leavevmode
\epsfxsize = .75 \columnwidth
\epsfbox{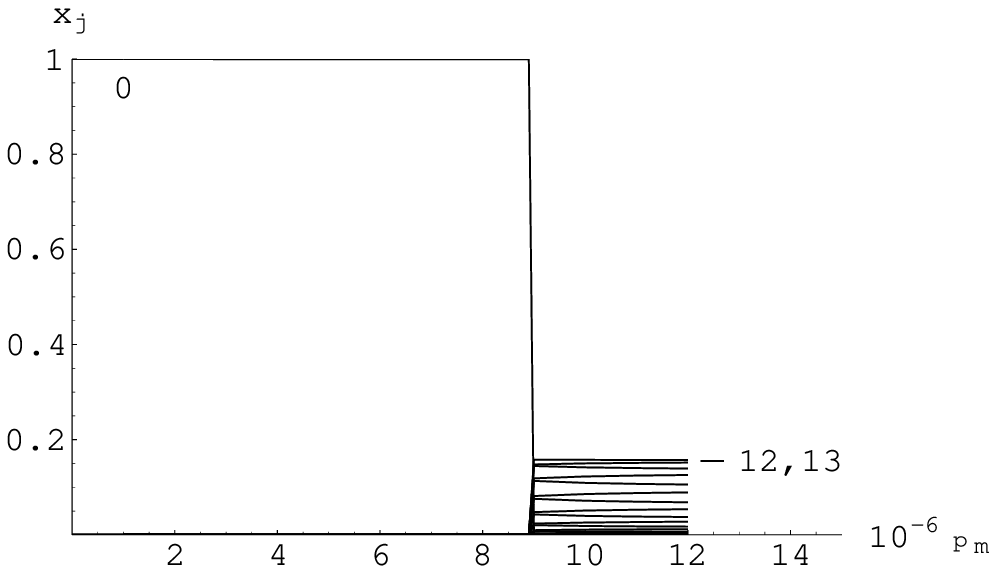}

\centering
\leavevmode
\epsfxsize = .75 \columnwidth
\epsfbox{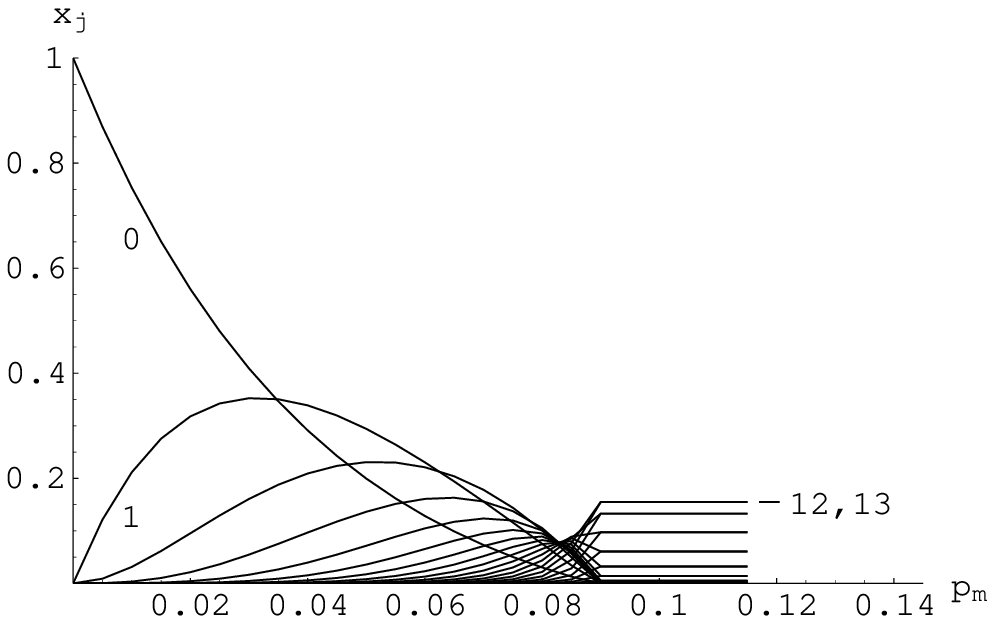}

\caption{The equilibrium distributions for recombination (upper graph) and pure
mutation (lower graph) dynamics, when the initial distribution is binomial
between the error classes. The gene sequences has length 25 and the fitness landscape
has an isolated peak ($A_0 = 10$, and $A_L = 1$ $\forall L \neq 0$).}
\label{numplot1}
\end{figure}

Fig.~\ref{numplot1} here.
\\

In this section we present results from computer simulations of the rate 
equations (\ref{eq8}). We concentrate on the asymptotic behavior as time goes
to infinity, and do not consider detailed  dynamics of the transients.
Equilibrium distributions are obtained  by a straight-forward simulation
 of the differential equations. All the simulations 
in this section  use uniform crossover ($p_c = 0.5$), 
which preserves the error class symmetry. 

We now consider a fitness landscape with an isolated peak ($A_0 = 10$, and $A_L = 1$ 
$\forall L \neq 0$). The equilibrium distributions for recombining and non-recombining populations 
are presented in figure~\ref{numplot1}, where the initial distribution is
binomial over the error classes. The phase transition between the localized and 
non-localized state is extremely sharp in the recombination case. The phase transition
occurs at a mutation rate which is orders of magnitude lower 
than in the non-recombining population.      

\begin{figure}[h]
\centering
\leavevmode
\epsfxsize = .75 \columnwidth
\epsfbox{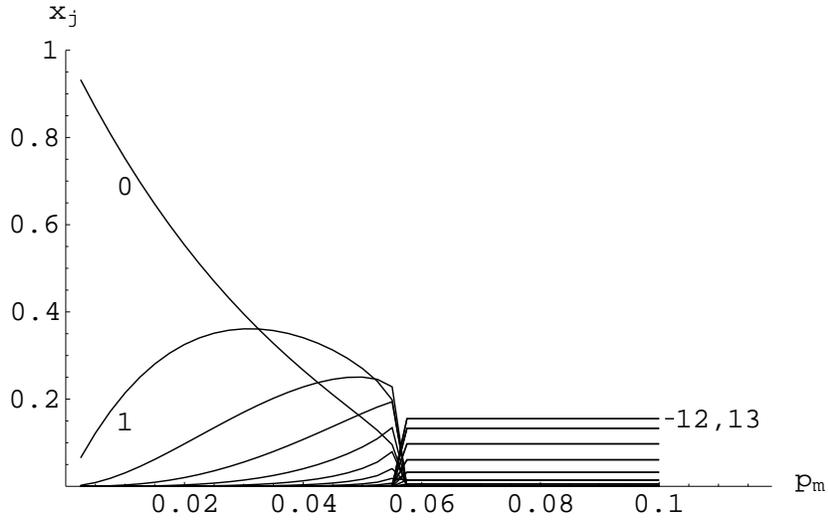}
\caption{The equilibrium distribution for a recombining population when
the initial distribution is concentrated to the master sequence, 
$x_0 = 1$, and $x_K = 0$ $\forall K \neq 0$.  The gene sequences have length 
25 and the fitness landscape has an isolated peak ($A_0 = 10$, and $A_L = 1$ 
$\forall L \neq 0$).}
\label{numplot2}
\end{figure}

Figure~\ref{numplot2} shows the equilibrium distribution of recombination dynamics 
with the same fitness landscape as  figure~\ref{numplot1}; the only difference 
is the initial distribution which is completely localized to the master sequence
($x_0 = 1$, and $x_K = 0$ $\forall K \neq 0$). We see that the equilibrium distributions
depend strongly on the initial distributions. The error threshold 
is still lower than in the pure mutation case, however the difference is
 much smaller. In general recombination
in single peak fitness landscapes tends to mix the gene sequences and push the
population above the error threshold.\\
\\
Fig.~\ref{numplot2} here.
\\

\begin{figure}[h]
\centering
\leavevmode
\epsfxsize = .75 \columnwidth
\epsfbox{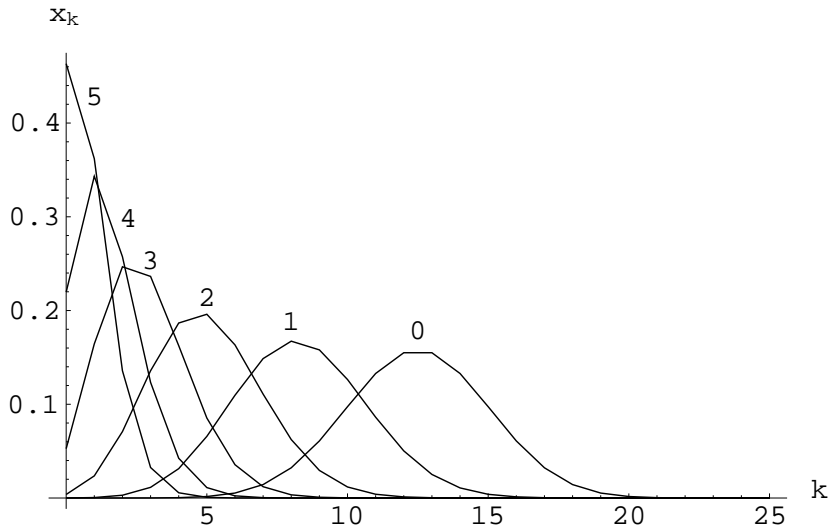}

\caption{Initial distributions for different values of the parameter $s $.}
\label{dist}
\end{figure}

\begin{figure}[h]
\centering
\leavevmode
\epsfxsize = .75 \columnwidth
\epsfbox{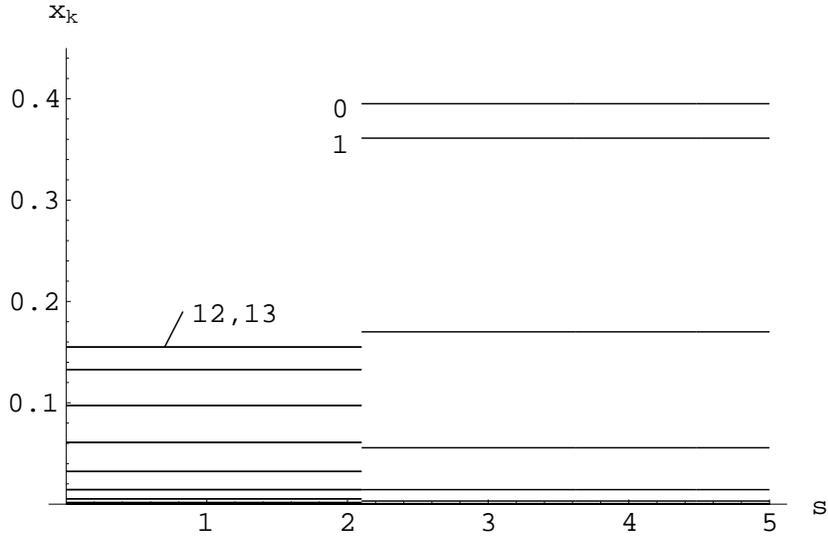}

\caption{Equilibrium distributions for different values of the parameter $s $. The copying fidelity
is constant $q = 0.97$. Note that there are only two different equilibrium distributions.}
\label{numplot25}
\end{figure}

\begin{figure}[h]
\centering
\leavevmode
\epsfxsize = .75 \columnwidth
\epsfbox{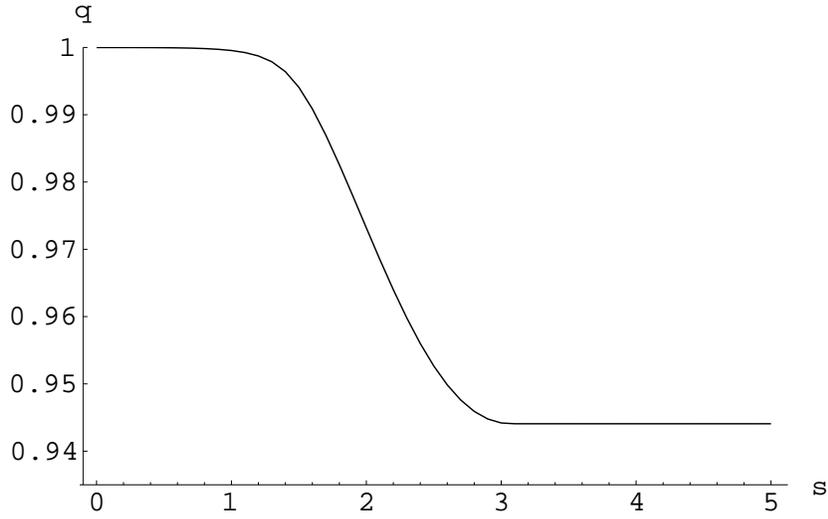}
\caption{The copying fidelity at the phase-transition for different initial distributions
$x_k (s )$ (as defined in equation~\ref{init}). The gene
sequence has length 25 and the fitness landscape has an isolated peak ($A_0 = 10$, and $A_L = 1$ 
$\forall L \neq 0$).}
\label{numplot3}
\end{figure}

Figure~\ref{numplot25} and~\ref{numplot3} show how the equilibrium distributions and the 
phase transition point varies with the initial distribution. The initial distributions are given by 

\begin{eqnarray}
x_k (s ) & = & \frac{ 2^{-s \cdot k} \left( \begin{array}{c} N \\ k \end{array} \right)}
                {\left( 1 + 2^{-s} \right) ^N}
\label{init}
\end{eqnarray} 
This gives a uniform distribution for $s =0$ and 
a distribution concentrated to the master-sequence for large $s$.
The graphs in figure~\ref{dist} show the initial 
distributions for some discrete 
 parameter values, $s = 0 , 1 , \cdots ,5$. Figure~\ref{numplot25} shows that there are two different
regions in the space of initial distributions, converging to two different fixed points.
In one corner of this space 
all the genomes are master-sequences. If the concentration vector starts out far from this corner
it will not converge into the corner unless the mutation rate is extremely low 
(as illustrated in figure~\ref{numplot1} 
or by the case of $s \in [ 0,1 ] $ in figure~\ref{numplot3}). If the initial distribution 
starts near the corner it will converge 
into the corner for much larger mutation rates (see figure~\ref{numplot2} or the region
$s \in [ 3 , 5 ] $ in 
figure~\ref{numplot3}). Figure~\ref{numplot3} shows the 
location of the phase transition point for different
initial distributions defined by equation~\ref{init}. This phase diagram shows how the border between the
two regions in figure~\ref{numplot25} changes with mutation rate. A change of $p_m$ from $9 \cdot 10^{-6}$ to
$0.055$, changes the border from $s =1$ to $3$. When the mutation rate is too low or too high only
one region exists corresponding to a single stable fixed-point. 

That there is an upper bound on the mutation rate where a stable localized fixed point ceases to exist
is obvious. The existence of a lower bound, below which all initial distributions converge to a 
localized distribution, is however non-trivial. This lower bound always exists and we will
present a proof of this in section~\ref{singlefix}.\\
\\
Fig.~\ref{dist} here.\\
Fig.~\ref{numplot25} here.\\
Fig.~\ref{numplot3} here.\\

The main conclusion to be drawn from these numerical simulations is that, for a wide
range of mutation rates, one finds a coexistence of two different equilibrium distributions
to the rate equations involving both recombination and point mutations. Which of these 
fixed points the population will converge to depends on the initial distribution. This means that
the space of initial distributions consists of two regions, with the border between this regions 
depending on the mutation rate. The whole range of mutation rates where a localized fixed point
exists is however lower than the phase transition point in the non-recombining case. This shows that
a recombining population is more sensitive to mutation than a non-recombining one on a
single peak landscape. Similar conclusions have been reached in a simpler model by
Bergman and Feldman~\cite{Feldman}. Similar results have also been shown in other work, 
see e.g.,~\cite{Boerlijst}.

\section{Existence of a single fixed-point at zero mutation rate.}

\label{singlefix}

In this section we investigate the behavior of the rate equations when $p_m \rightarrow 0^+$.
In section~\ref{num} it was shown numerically that at very low mutation rates, all initial distributions converge 
to a highly localized equilibrium distribution. Here we show that this region always 
exists for fitness landscapes fulfilling certain assumptions, to be specified below.

The idea behind the proof is to study the dynamics of one position or loci in the genome and sum
over all possibilities at the other positions. Let $S ^{ ( N, n , i)}_{\alpha}$ denote all genomes of length
$N$ that contain the sequence $\alpha$ starting at position $1 \leq i \leq N-n$, where $\alpha$ is an index coding for genomes 
of length $n$. For 
example; $S^{ ( 10,2,1 )}_3$ will be all genomes of length $10$ that starts with $( 1 1 )$. We also 
introduce the notation $x^{(N)}_k$, where $N$ simply indicates the genome length and 
affects decoding of the index $k$. We can now write the rate equations~(\ref{eq8}) as

\begin{eqnarray}
        \dot{x}^{(N)}_k & = & \sum _{l , m} V_k^{(N) l m} A_l x^{(N)}_l A_m x^{(N)}_m - 
                               \left( \sum _l A_l x_l ^{(N)} \right) ^2 x^{(N)}_k  
\end{eqnarray}
The crossover operator has the following property

\begin{eqnarray}
        \sum _{k \in S^{(N,n,i)}_{\alpha} } T_k ^{l m} & = & T_{\alpha}^{\beta \gamma} \mbox{ for } l \in S^{(N,n,i)}_{\beta} , 
                                                         m \in S^{(N,n,i)}_{\gamma}, \forall i
\end{eqnarray}
where no assumptions on the crossover probability in made.

Since the point mutation operator $Q^{(N) l}_k$ has the same property, so will the combined operator
$V^{(N) l m}_k$. We can now use this property and sum the rate equations over all sequences in $S^{(N, 1,i)}_{\alpha}$

\begin{eqnarray}
        \sum _{k \in S^{(N , 1,i)}_{\alpha}} \dot{x}^{(N)}_k & = & \sum _{k \in S^{(N , 1,i)}_{\alpha}} \left(
                    \sum_{l , m} V_k^{(N) l m} A_l x^{(N)}_l A_m x^{(N)}_m \right. \\ 
                  & & \left. -  \left( \sum _l A_l x_l ^{(N)} \right) ^2 x^{(N)}_k \right) \Rightarrow \nonumber \\
        \dot{x}^{(1)}_{\alpha} & = & \sum _{\beta , \gamma} V_{\alpha}^{(1) \beta \gamma} \sum _{l \in S^{(N , 1,i)}_{\beta}} A_l x^{(N)} _l 
            \sum _{m \in S^{(N , 1,i)}_{\gamma}} A_m x^{(N)} _m \\
                     & & - \left( \sum _{\beta} \sum _{l \in S^{(N , 1,i)}_{\beta}} A_l 
            x^{(N)} _l \right) ^2 x_{\alpha}^{(1)} \label{one_loci_eq}
\end{eqnarray}
The following, compact, notation is now introduced:

\begin{eqnarray}
        \sum _{ l \in S_{\beta}^{(N , 1,i)}} A_l x^{(N)}_l & = & \left\{ \begin{array}{lcl}  \Delta ^{(i)}_0 & \mbox{ if } & \beta = 0 \\
                                   \Delta ^{(i)} _1 & \mbox{ if } & \beta = 1 \end{array} \right.
\end{eqnarray}
Eq.~\ref{one_loci_eq} simplifies to

\begin{eqnarray}
        \dot{x}_0^{(1)} & = & q \left( \Delta ^{(i)} _0 \right) ^2 + \Delta ^{(i)} _0 \Delta ^{(i)} _1 + (1-q) \left( \Delta ^{(i)} _1 \right) ^2 -  
		\left( \Delta ^{(i)} _0 + \Delta ^{(i)} _1 \right) ^2 x_0 ^{(1)} \\
        x_1^{(1)} & = & 1 - x_0 ^{(1)}
\end{eqnarray}
which, in the limit $q \rightarrow 1^-$, simplify to

\begin{eqnarray}
        \dot{x}_0 ^{(1)} & = & \left( \Delta ^{(i)} _0 + \Delta ^{(i)} _1 \right) \left( \Delta ^{(i)} _0 x_1^{(1)} - \Delta ^{(i)} _1 x_0 ^{(1)} \right) \nonumber \\
	x_1 ^{(1)} & = & 1 - x_0 ^{(1)} 
	\label{part_result}
\end{eqnarray}
To continue the following assumption on the fitness landscape is needed:

\begin{eqnarray}
	A _l & \leq & A_m \hspace{0.3cm} \mbox{if} \hspace{0.1cm} l \in S ^{N,1,i} _1 ,  m \in S ^{N,1,i} _0
\label{assumption1}
\end{eqnarray}
We further assume that there exist a gene sequence $M \in S ^{N,1,i} _0$ such that

\begin{eqnarray}
	A _l & < & A_M \hspace{0.2cm} \forall l \in S ^{N,1,i} _1
\label{assumption2}
\end{eqnarray}
These two assumptions mean that no sequences with a zero at position $i$ have a fitness 
inferior to any sequence with a one  at this position, and that there exist
at least one sequence with with a zero at position $i$ with strictly larger
fitness than the sequences with a one at this position. Under these assumptions, 
the following inequalities hold

\begin{eqnarray}
	\Delta ^{(i)} _0 & \geq & \Delta ^{(i)} _{0, min} x^{(1)} _0 \nonumber \\
	\Delta ^{(i)} _1 & \leq & \Delta ^{(i)} _{1, max} x^{(1)} _1 
\label{estimate}
\end{eqnarray}
where $\Delta ^{(i)} _{0, min}$ ($ \Delta ^{(i)} _{1, max}$) denotes the minimum (maximum)
fitness of the sequences with a $0$ ($1$) at position $i$. We further note that at least one of the
inequalities in Eq.~\ref{estimate} is strict unless $x _{M} ^{(N)} =0$ for all $M$ fulfilling 
Eq.~\ref{assumption2}. Eq.~\ref{estimate} implies the following estimate

\begin{eqnarray}
	\dot{x}_0 ^{(1)} & \geq & \left( \Delta ^{(i)} _{0, min} -
		\Delta ^{(i)} _{1, max} \right) x^{(1)} _1 x^{(1)} _0
\label{result}
\end{eqnarray}
with equality if and only if  $x _{M} ^{(N)} =0$ for all $M$ fulfilling 
Eq.~\ref{assumption2} or $ x^{(1)} _1 =0$ or $x^{(1)} _0 =0$. Note however 
that $x_0^{(1)}=0$, $x_1^{(1)}=1$ is an (unstable) fixed-point since no mutations 
implies no inventions of new genes.

From Eq.~\ref{result} it is clear that the rate equations will converge to a state
where all sequences has a zero at position $i$.
  This fixed point is unstable and it is clear that they cease to exist when
the mutation rate is non-zero.

We conclude that all sequences with a one at position $i$ will diminish 
after long time, and can therefore be be discarded. We can then search for 
a new position such that the remaining half of the fitness landscape 
satisfies the assumptions in 
Eq.~\ref{assumption1} and~\ref{assumption2}. If this can be repeated (possibly interchanging 
the zero and one as being superior, since this choice is arbitrary) until the last
position, we conclude that the rate equations converge to a state completely
dominated by genomes with the same sequence (which necessarily is a global optimum).
Loosely, we may describe such fitness landscapes as having a natural ordering of the
importance of its loci. One example of a fitness landscape fulfilling these requirements
is a single peaked fitness landscape, describing a degenerate case where the 
positions can be chosen arbitrarily.

\section{Conclusions and discussion}
\label{discussion}

We have studied Eigen's quasispecies model extended
to include crossover as well as mutations.
The numerical simulations of section~\ref{num} show that there are significant changes 
in the dynamics of the rate equations because of the non-linearity arising from
the introduction of crossover. For a wide range of mutation rates, 
two simultaneous stable fixed points
exist. One fixed point is concentrated around the master sequence while the other describes 
a uniform distribution. For extremely low and rather high mutation frequencies 
there is only
a single fixed point, corresponding to the localized distribution and the 
uniform one, respectively.
The mutation frequency at the point where the localized fixed point ceases to 
exist is still lower than the error threshold without recombination.

In this paper we prove that, for a class of fitness landscapes having a hierarchical 
ordering of the loci in the genome (see Section~\ref{singlefix} for details),
a single globally stabile fixed point exist in the limit of zero mutation rate.
Since the proof is valid for all crossover probabilities, the only natural 
generalization is to expand the class of fitness landscapes. A possible
generalization of the technique in Section~\ref{singlefix} could be to prove that; 
within larger class of 
fitness landscapes, for any point in time, i.e., for any distribution $\vec{x}^{(N)}$,
we can always find a position $i$ such that Eq.~\ref{result} is fulfilled.
The position $i$ would now depend on the distribution (which changes in time), 
not only the fitness landscape which is the case in our proof. Technically however,
this generalization is non-trivial since the changing of position with the 
distribution makes it complicated to argue that all locus in the global fixedpoint
will dominate completely in the infinite time limit.

\bibliographystyle{unsrt}

\bibliography{evolution}

\begin{thebibliography}{10}

\bibitem{Eigen71}
M.~Eigen.
\newblock Self-organization of matter and the evolution of biological
  macromolecules.
\newblock {\em Naturwissenschaften}, 58:465--523, 1971.

\bibitem{Eigen77}
M.~Eigen and P.~Schuster.
\newblock The hypercycle. {A} principle of natural self-organization. {P}art
  {A}: emergence of the hypercycle.
\newblock {\em Naturwissenschaften}, 64:541--565, 1977.

\bibitem{Schuster86}
P.~Schuster.
\newblock Dynamics of molecular evolution.
\newblock {\em Physica D: Nonlinear Phenomena}, 16:100--119, 1986.

\bibitem{Schuster85}
P.~Schuster and K.~Sigmund.
\newblock Dynamics of evolutionary optimization.
\newblock {\em Berichte der Bunsen-Gesellschaft, Physical Chemistry},
  89:668--682, 1985.

\bibitem{Swetina88}
J.~Swetina and P.~Schuster.
\newblock Stationary mutant distribution and {E}volutionary {O}ptimization.
\newblock {\em Bulletin of Mathematical Biology}, 50:635--660, 1988.

\bibitem{Leuthausser86}
I.~Leuth\"ausser.
\newblock An exact correspondence between {E}igen's evolution model and a
  two-dimensional {I}sing system.
\newblock {\em J. Chem. Phys.}, 84(3):1884--1885, 1986.

\bibitem{Tarazona92}
P.~Tarazona.
\newblock Error thresholds for molecular quasispecies as phase transitions:
  {F}rom simple landscapes to spin-glass models.
\newblock {\em Physical Review A}, 45(8):6038--6050, 1992.

\bibitem{AF98}
D.~Alves and J.F. Fontanari.
\newblock Error thresholds in finite populations.
\newblock {\em Physical Review E.}, 57(6):7008--7013, 1998.

\bibitem{Wright}
S.~Wright.
\newblock The roles of mutations, inbreeding, crossbreeding and selection in
  evolution.
\newblock {\em Proceeding of the Sixth International Congress on Genetics},
  1:356--366, 1932.

\bibitem{Kauffman87}
S.A. Kauffman and S.~Levin.
\newblock Towards a general theory of adaptive walks on rugged landscapes.
\newblock {\em Journal of Theoretical Biology}, 128:11--45, 1987.

\bibitem{Palmer91}
R.~Palmer.
\newblock Optimization on rugged landscapes.
\newblock In A.S. Perelson and S.A. Kauffman, editors, {\em Molecular Evolution
  on Rugged Landscapes: Proteins, RNA and the Immune System}, pages 3--25,
  Redwood City, CA, 1991. Addison Wesley.

\bibitem{Fontana93}
W.~Fontana, P.F. Stadler, E.G. Bornberg-Bauer, T.~Griesmacher, I.L. Hofacker,
  M.~Tacker, P.~Tarazona, E.D. Weinberger, and P.~Schuster.
\newblock {RNA} folding and combinatory spaces.
\newblock {\em Physical Review E}, 47:2083--2089, 1993.

\bibitem{Macken91}
P.~S. Hagan, C.~A. Macken, and A.~S. Perelson.
\newblock Evolutionary walks on rugged landscapes.
\newblock {\em SIAM Journal of Applied Mathematics}, 51:799--827, 1991.

\bibitem{Stadler95a}
P.F. Stadler.
\newblock Landscapes and their correlation function.
\newblock {\em Journal of Mathematical Chemistry}, 20:1--45, 1996.

\bibitem{Boerlijst}
S.~Bonhoeffer M.C.~Boerlijst and M.A. Nowak.
\newblock Viral quasi-species and recombination.
\newblock {\em Proceedings of the Royal Society of London B}, 263:1577--1584,
  1996.

\bibitem{OH98}
G.~Ochoa and G.~Harvey.
\newblock Recombination and error thresholds in finite populations.
\newblock In W.~Banzhaf and C.~Reeves, editors, {\em Foundations of Genetic
  Algorithms (FOGA-5)}, San Francisco:CA, 1998. Morgan Kaufmann.
\newblock ftp://ftp.cogs.susx.ac.uk/pub/users/inmanh/fogat.ps.gz.

\bibitem{Stadler96}
P.F. Stadler and G.P. Wagner.
\newblock The algebraic theory of recombination spaces.
\newblock {\em Evolutionary Computation}, 5:241--275, 1997.

\bibitem{Feldman}
A.~Bergman and M.W. Feldman.
\newblock Recombination dynamics and the fitness landscape.
\newblock {\em Physica D}, 56:57--67, 1992.

\bibitem{Monroe}
S.~Monroe and M.~Schlesinger.
\newblock {\em Proceedings of National Academy of Science, USA}, 80:3279--3283,
  1983.

\bibitem{Li}
T.~Li \and J.Y.~Zhang.
\newblock {\em Journal of Virology}, 74(16):7646--7650, 2000.

\bibitem{Holland75}
J.~Holland.
\newblock {\em Adaptation In Natural and Artificial Systems}.
\newblock The University of Michigan Press, 1975.

\bibitem{Maynard70}
J.~Maynard~Smith.
\newblock Natural selection and the concept of protein space.
\newblock {\em Nature}, 225:563--564, 1970.

\bibitem{Jones}
B.L. Jones, R.H. Enns, and S.S. Rangnekar.
\newblock On the theory of selection in coupled macromolecular systems.
\newblock {\em Bulletin of Mathematical Biology}, 38:15--28, 1976.

\bibitem{Thomson}
C.J. Thomson and J.L. McBride.
\newblock On {E}igen's theory of self-organization of matter and the evolution
  of biological macromolecules.
\newblock {\em Mathematical Bioscience}, 21:127--142, 1974.

\bibitem{Bellman}
R.~Bellman.
\newblock {\em Introduction to {M}atrix {A}nalysis}.
\newblock McGraw-Hill, New York, 1970.

\end{thebibliography}

\newpage

\newpage

\newpage

\newpage

\newpage

\newpage

\newpage

\end{document}